# An integral method for the calculation of the reduction in interfacial free energy due to interfacial segregation

Ivan Blum[1]†, Sung-Il Baik[1], Mercouri G. Kanatzidis[2], and David N. Seidman[1,3,*]

[1]Department of Materials Science and Engineering, Northwestern University, Evanston, Illinois 60208-3108, USA

[2]Department of Chemistry, Northwestern University, Evanston, Illinois 60208-3113

[3]Northwestern University Center for Atom-Probe Tomography, Evanston, IL 60208-3108 USA

**Abstract**

A method based on the Gibbs' adsorption isotherm is developed to calculate the decrease in interfacial free energy resulting from solute segregation at an internal interface, built on measured concentration profiles. Utilizing atom-probe tomography (APT), we first measure a concentration profile of the relative interfacial excess of solute atoms across an interface. To accomplish this we utilize a new method based on J. W. Cahn's formalism for the calculation of the Gibbs interfacial excess. We also introduce a method to calculate the decrease in interfacial free energy that is caused by the segregating solute atoms. This method yields a discrete profile of the decrease in interfacial free energies, which takes into account the measured concentration profile and calculated Gibbsian excess profile. We demonstrate that this method can be used for both homo- and hetero-phase interfaces and takes into account the actual distribution of solute atoms across an interface as determined by APT. It is applied to the case of the semiconducting system PbTe-PbS 12 mol.%-Na 1 mol.%, where Na segregation at the PbS/PbTe interface is anticipated to reduce the interfacial free energy of the {100} facets. We also consider the case of the nickel-based Alloy 600, where B and Si segregation are suspected to impede inter-granular stress corrosion cracking (IGSCC) at homo- (GB) and hetero-phase metal carbide ($M_7C_3$) interfaces. The concentration profiles associated with internal interfaces are measured by APT using an ultraviolet (wavelength = 355 nm) laser to dissect nanotips on an atom-by-atom and atomic plane-by-plane basis. (248 words)

**Corresponding author:** David N Seidman, *E-mail address:* d-seidman@northwestern.edu

†Current institution : GPM UMR 6634, Université de Rouen, Avenue de l'Université, BP12,

76801 Saint Etienne du Rouvray, France

1. **Introduction**

Interfacial segregation can have important effects on the microscopic properties of internal interfaces and eventually on the macroscopic properties of a bulk material, such as mechanical, electrical or chemical properties [1,2]. The driving force for interfacial segregation is the reduction of the interfacial free energy. This change in interfacial free energy can affect the behavior of the interface. It can, for instance, impede grain growth by decreasing the interfacial free energy of grain boundaries [3,4], or in the case of matrix/precipitates interfaces the reduction in interfacial free energy can decrease the coarsening rate of the precipitates [5–7]. Therefore, there is a strong interest in determining the decrease in interfacial free energy caused by segregation. Different approaches to this problem exist in the literature [8–11]. The decrease in the interfacial free energy is related to the Gibbsian interfacial excess of solute, which is the excess number of atoms segregated at an internal interface per unit area of interface [12,13], which is measureable for both homophase [13,14] and heterophase [15–17] interfaces by atom-probe tomography (APT). The calculation of the change of the interfacial free energy requires the integration of the solute's chemical potential from a theoretical initial concentration at the interface, without segregation, to the final concentration in the presence of segregation. Very often, in practice, the exact relationship between the chemical potential of an atom and the solute concentration is not known and the calculation of the reduction in interfacial energy can only be done within Henry's law approximation, that is, that the solute concentration is small enough for the solute's activity coefficient to be independent of the concentration. In this case, only the initial and final concentrations need to be known; i.e., the concentrations in the absence of segregation, and the concentrations at local equilibrium. How to obtain this information isn't necessarily obvious, however. According to Gibbs' description of an excess quantity at an interface, both phases are extended to the interface, which is considered to be a simple geometric dividing surface. For a heterophase interface, across which a solute partitions, which is a very common case, the initial concentration in the absence of any segregation is, therefore, undetermined. In practice, for the calculation of the decrease in the interfacial free energy, this concentration is commonly chosen to be equal to the matrix solute concentration [8–11], which is an arbitrary choice that doesn't have a thermodynamic basis. And the final concentration in the presence of segregation is chosen to be the maximum concentration at the interface, which neglects the shape of the interface profile.

In this article, we describe a different method for the calculation of the reduction of the interfacial free energy caused by segregation at an internal interface. This method is established using Cahn's interface description, which does not utilize a dividing surface [18]. We first define the interfacial excess according to Cahn and next the methodology for the calculation of the decrease in interfacial free energy. We then apply this methodology to two different systems: firstly, the semiconducting system PbTe-PbS 12 mol.%-Na 2 mol.%, which is studied for its thermoelectric properties and in which Na segregation at the PbS/PbTe interfaces was observed to affect the PbS precipitates' morphologies [19–21]. Increasing the Na concentration at the interfaces increases the dominance of {100} facets compared to {111} facets, showing that Na reduces the interfacial free energy of the {100} facets. Secondly, we use this methodology to study segregation at a metal carbide at a grain boundary in the Ni-based 600 alloy, in which segregation is expected to reduce intergranular stress induced corrosion cracking (IGSCC) [22,23]. We demonstrate that compared to the methodology that is widely used, the present methodology has a quantitative thermodynamic significance and can be applied equally well to homo- and hetero-phase interfaces.

2. Gibbsian interfacial excess

The Gibbsian interfacial excess of an element, A, is the excess number of solute atoms per unit area segregated at an interface between two phases, $\Gamma_A$, compared to the number of solute atoms already present in the two phases where interfacial segregation is absent. To avoid redundancy, in the remainder of this article, all extensive quantities are implicitly expressed per unit area of interface present in the considered volume. The Gibbsian interfacial excess is defined as:

$$\Gamma_A = N_A^m - N_A^0 \qquad (1)$$

where $N_A^m$ is the number of A atoms present in the volume and $N_A^0$ is the reference number of A atoms, which would be expected in the same volume in the absence of interfacial segregation. In the absence of interfacial segregation, the total composition of the volume containing the interface can be expressed as a linear mixture of the phases labeled α and β. Therefore, one defines this reference number of atoms as:

$$N_A^0 = N^\alpha X_A^\alpha + N^\beta X_A^\beta ; \qquad (2)$$

where $N^i$ is the number of atoms belonging to phase i (i = α or β) and $X_A^i$ is the atomic fraction of element A in phase i. While the concentrations in the two phases, $X_A^\alpha$ and $X_A^\beta$, are measurable by different techniques, the total number of atoms belonging to each one of the two phases in the reference volume represents two degrees of freedom that depend on how this volume is defined. The different methods to calculate the interfacial excess use different information to remove these two degrees of freedom. In Gibbs' method for calculating the absolute interfacial excess, a dividing surface is placed at the interface, whose position is arbitrarily chosen [24].

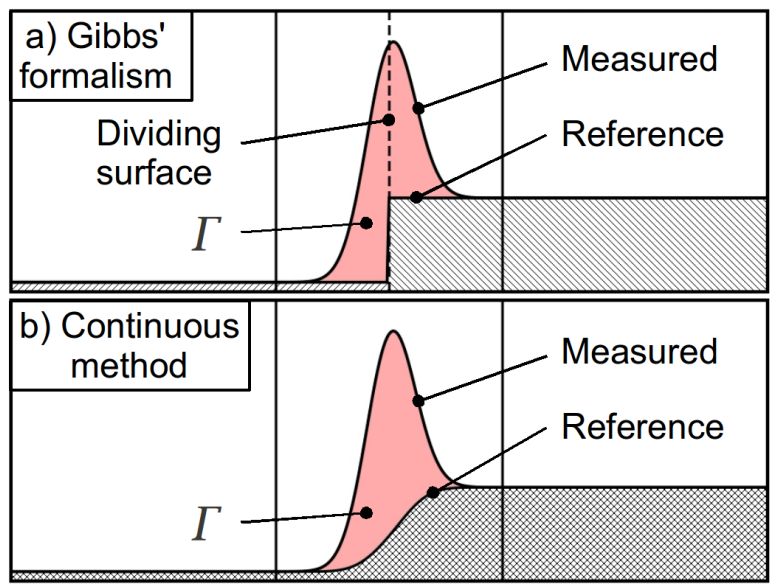

**Figure 1.** Schematic description of (a) Gibbs's method for the calculation of the interfacial excess, using a dividing surface, and (b) the continuous method presented in this article, which is derived from Cahn's formulation for measuring the interfacial excess. The continuous reference concentration profile permits the calculation of a profile of the decrease in interfacial free energy.

As illustrated, Fig. 1(a), defining the position of the interface defines the amount of the two phases in the interfacial region and thereby defines the reference amount of solute present at the interface in the absence of segregation (gray area). The resulting interfacial excess, the red area, depends strongly on the position chosen for the interface and therefore, doesn't have a true thermodynamic meaning. Gibbs [25] ingeniously solved this problem by defining the relative interfacial excess, with respect to one or two other elements whose excesses are relatively small. For this case the relative excess of A with respect to one element, B, is given by:

$$\Gamma_A^B = \Gamma_A - \Gamma_B \frac{\left(X_A^\alpha - X_A^\beta\right)}{\left(X_B^\alpha - X_B^\beta\right)} \tag{3}$$

And the relative excess of A with respect to two elements, B and C, is given by:

$$\Gamma_A^{B,C} = \Gamma_A - \frac{\Gamma_C\left(X_A^\alpha X_B^\beta - X_B^\alpha X_A^\beta\right) - \Gamma_B\left(X_A^\alpha X_C^\beta - X_C^\alpha X_A^\beta\right)}{\left(X_C^\alpha X_B^\beta - X_B^\alpha X_C^\beta\right)} \tag{4}$$

The result for the relative Gibbsian interfacial excess does not depend on the position of the interface, but the calculation of the absolute excesses does require the positioning of an interface [12,25,26]. More elegant solutions that do not require the positioning of an interface are given by Cahn [18] and later Umantsev [27] for two and one reference elements, respectively. For one reference element, one obtains the following system of equations:

$$N_A^0 = N^\alpha X_A^\alpha + N^\beta X_A^\beta \tag{5.a}$$

$$N_B = N^\alpha X_B^\alpha + N^\beta X_B^\beta ; \tag{5.b}$$

$$N^{tot} = N^\alpha + N^\beta ; \tag{5.c}$$

where $N^{tot}$ is the total number of atoms between the two dividing surfaces. Solving this system of equations yields an expression for $N_A^0$, and substituting it into Eq. (1) yields the following equation for the interfacial excess of A with respect to B :

$$\Gamma_A^B = N^{tot}\left[X_A^m - X_A^\beta - \left(X_B - X_B^\beta\right)\frac{\left(X_A^\alpha - X_A^\beta\right)}{\left(X_B^\alpha - X_B^\beta\right)}\right]; \tag{6}$$

which is equivalent to Umantsev's expression for the interfacial excess that he expressed employing atomic densities [27]. For two reference elements, Eq. (5.c) is replaced by the expression for the amount of a second reference element C:

$$N_C = N^\alpha X_C^\alpha + N^\beta X_C^\beta. \tag{7}$$

Replacing Eq. (5.c) with Eq. (7) in the system of equations (5) yields the following solution for $N_A^0$:

$$N_A^0 = \frac{N_C\left(X_A^\alpha X_B^\beta - X_B^\alpha X_A^\beta\right) - N_B\left(X_A^\alpha X_C^\beta - X_C^\alpha X_A^\beta\right)}{\left(X_C^\alpha X_B^\beta - X_B^\alpha X_C^\beta\right)} \tag{8}$$

which, combined with Eq. (1) yields the interfacial excess of A with respect to elements B and C, according to Cahn's formalism [18]:

$$\Gamma_A^{B,C} = N_A^m - \frac{N_C\left(X_A^\alpha X_B^\beta - X_B^\alpha X_A^\beta\right) - N_B\left(X_A^\alpha X_C^\beta - X_C^\alpha X_A^\beta\right)}{\left(X_C^\alpha X_B^\beta - X_B^\alpha X_C^\beta\right)} \tag{9}$$

Interestingly, developing Gibbs' two equations (Eq. (3) and (4)), for one and two reference elements, using the definition of the interfacial excess given by Eqs. (1) and (2) yields Umantsev's and Cahn's equations, respectively, demonstrating that they are mathematically equivalent to Gibbs' expressions. This is an important result because it demonstrates that there isn't an advantage in using Gibb's equations when compared to Cahn's and Umantsev's expressions, which are more straightforward and convenient, contrary to what was done in numerous prior publications on this subject. Another treatment was also developed by C. Wagner [28] and was later extended to line and point defects by Kirchheim [29,30].

## 3. Calculation of the decrease of the interfacial free energy

### 3.1 Prior approaches

The macroscopic driving force for segregation is the reduction of the interfacial free energy. The change in interfacial free energy is related to the Gibbsian interfacial excess by Gibbs' adsorption isotherm [31,32]:

$$\left(\frac{\partial \sigma}{\partial \mu_i}\right)_{T,P,\mu_1,\ldots,\mu_i-1,\mu_i+1,\ldots,\mu_n} = -\Gamma_i. \tag{10}$$

where $\sigma$ is the interfacial free energy, T and P are the temperature and pressure, $\mu_i$ and $\Gamma_i$ are the chemical potentials and the equilibrium Gibbsian interfacial excesses of specie $i$, respectively. In Henry's law (for an ideal solution or small concentrations), the activity coefficient is assumed to be a constant, which yields:

$$d\mu_i = k_B T \left(\frac{dC_i}{C_i}\right). \tag{11}$$

Integrating over the range of the change in concentrations, $C_i$, due to segregation yields the change in the interfacial free energy caused by segregation:

$$\Delta\sigma_i = -\Gamma_i^{relative} k_B T \cdot \ln\left(\frac{C_{final}}{C_{initial}}\right); \tag{12}$$

where $C_{initial}$ is the solute concentration at the interface in the absence of segregation and $C_{final}$ is the measured solute concentration in presence of segregation. According to Gibbs' description of an interface, if the solute has a higher concentration in one of the two phases, the solute concentration will be undetermined on the dividing surface, as shown in Fig. 1(a). Generally, $C_{initial}$ is set equal to the matrix concentration [8–11]. But the latter choice is arbitrary and doesn't have a true thermodynamic meaning. Choosing the matrix or precipitate concentrations as $C_{initial}$ can lead to different results. Moreover, $C_{final}$ is defined as the maximum solute concentration at the interface. This choice makes the assumption that all the segregating atoms are located in an interfacial region with a certain thickness having a homogeneous concentration equal to $C_{final}$, and it neglects the actual concentration profile measured across the internal interface.

### 3.2 New approach utilizing Cahn's description of an interface

These problems are inherent to Gibbs' formalism. A solution can, however, be found in Cahn's description of an interface [18]. As stated, the average composition of a volume containing a heterophase interface without segregation can be expressed as a mixture of the two pure phases.

This definition is also true for every point or bin located in the interface's concentration profile. Hence, the development that yielded equation (8), can also be performed for local concentrations measured at an interface. Thus, one can calculate a local reference concentration that would be obtained in the absence of segregation:

$$C_A^0(x) = \frac{C_C(x)(X_A^\alpha X_B^\beta - X_B^\alpha X_A^\beta) - C_B(x)(X_A^\alpha X_C^\beta - X_C^\alpha X_A^\beta)}{(X_C^\alpha X_B^\beta - X_B^\alpha X_C^\beta)} \quad (13)$$

where $C_i$ is the concentration of *specie i*. The resulting reference concentration profile is schematically represented in Fig. 1(b). We can also calculate the corresponding local excess using:

$$d\Gamma_i = (C_i^m(x) - C_i^0(x))dx \quad (14)$$

where $C_i^m$ is the local measured *concentration of i*. Using this definition, a local decrease in interfacial free energy can be calculated using an expression similar to eqn. (12):

$$d\Delta\sigma_i = -d\Gamma_i k_B T \cdot \ln\left(\frac{C_i^m(x)}{C_i^0(x)}\right) \quad (15)$$

where $d\Delta\sigma_i$ is the local change in interfacial free energy due to *specie i*. The total decrease in interfacial free energy is then obtained by integrating this function over the interval containing the interface to obtain an integral value of the interfacial free energy. When discretized it can be applied to a measured concentration profile, which yields:

$$\Delta\sigma_i = \sum_{n=1}^{N} -(C_i^m(n) - C_i^0(n))\Delta x k_B T \cdot \ln\left(\frac{C_i^m(n)}{C_i^0(n)}\right) \quad (16)$$

where n is the index of the bins of the concentration profile, N is the total number of bins, and $\Delta x$ is the bin width of the concentration profile. With this definition, the problem of the choice of the values for $C_{initial}$ and $C_{final}$ is alleviated because they are replaced by $C_i^0$ and $C_i^m$, respectively, which are well defined for every position in a concentration profile.

## 4. Experimental measurements

### 4.1 Atom-Probe Tomography (APT) and sample preparation

APT permits the analysis of a small volume of material in three-dimensions (3D) [33–35] often at a sub-nanometer scale [36]. The sample needs to have a needle-shaped geometry with a radius of curvature of ~40 nm at the nanotip's apex. The sample is subjected to a high electric field and is field evaporated on a nearly atom-by-atom and atomic plane-by-plane basis. Measuring the time- of-flight and impact position of the ions on a two-dimensional detector, one then reconstructs the evaporated volume in 3D. For conducting materials, the evaporation can be triggered by a voltage pulse, whereas for poorly conducting materials one uses a laser pulse applied to the specimen [37,38], permitting the analysis of semiconducting [39–41] or dielectric materials [42–45]. APT has proven to be useful for the characterization of bulk materials and interfaces in a wide range of materials systems. In this research, APT was utilized to characterize interfaces and the method developed in this article was applied to measured segregation profiles for three different materials systems.

The APT specimen preparation utilized a focused ion-beam (FIB) microscope, which enables site specific sampling of specific interfacial regions and sample sharpening to an end radius of <100 nm [46,47]. In the FIB sample preparation method, a wedge-shaped piece (30×5 μm$^2$) of material is extracted from a sample using a Ga$^+$ ion-beam, lifted-out, and cut into different sections, which are mounted and welded onto Si microposts using ion-beam assisted Pt deposition. Finally, a sharp nanotip is formed by annular milling with a radius of curvature of less than ~40 nm, which is required for an APT experiment. In the case of the metallic system, the sample was first analyzed using electron backscattering diffraction (EBSD) prior to the FIB microscope lift-out process, in order to determine a GB's macroscopic five degree of freedom [24,48,49]. Then a wedge was lifted-out from selected regions containing grain boundaries or heterophase interfaces and attached to an axial rotating-micromanipulator [50]. Then the wedge-shaped piece was rotated to place the GB plane in a horizontal orientation, and subsequently welded to a Si micro-post.

### 4.2 Analysis of a semiconducting system (PbTe-PbS)

The methodology presented in this article is first applied to the semiconductor PbTe-PbS 12 mol.%-Na 1 mol.%, which is studied for its thermoelectric properties [19–21]; it contains PbS precipitates that improve the thermoelectric figure of merit by increasing the scattering of long wavelength phonons. The sample was prepared by melting the elements at 1373 K, which was followed by slowly air cooling the alloy during ~3 h to room temperature: more details concerning sample preparation can be found elsewhere [20]. It was analyzed by APT at a temperature of 32 K, utilizing a picosecond ultraviolet laser (wavelength = 355 nm) pulsed at 250 kHz using 30 pJ per pulse. The evaporation rate was fixed at 0.005 atom per pulse (0.5 %). Under these conditions, the charge state ratio of $Pb^{++}/Pb^{+}$ was equal to 4.78 for the PbTe matrix and 3.85 for the PbS phase. Data was then reconstructed in three-dimensions and analyzed utilizing Cameca's IVAS 3.6 software package.

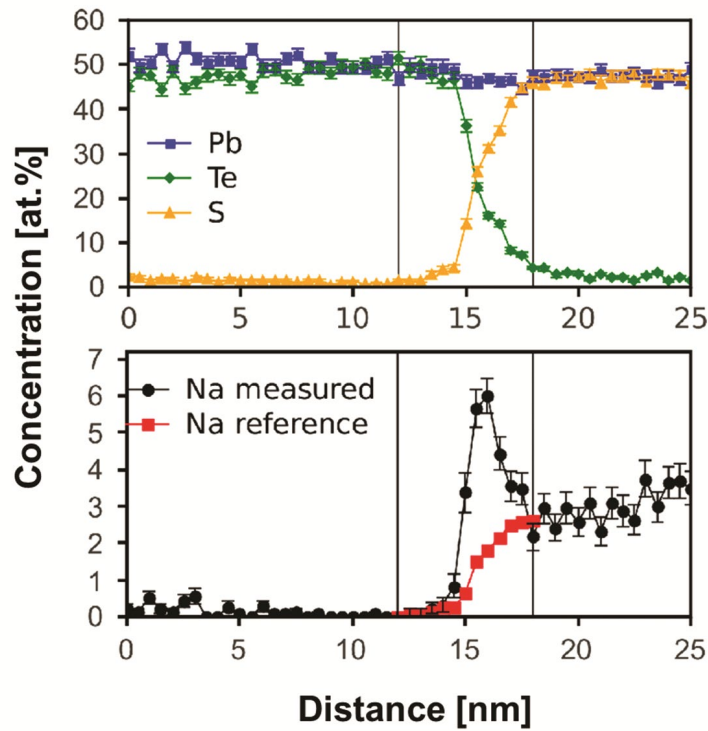

**Figure 2.** Concentration profiles of Pb, Te, S and Na measured, using atom-probe tomography, at the interface between the PbTe matrix and a PbS precipitate. The reference Na concentration profile is calculated using Te and S as reference elements.

Fig. 2 displays concentration profiles measured across a (100) facet of a PbS precipitate in the case of 1 mol.% Na doping, using an analysis cylinder of 14.3 nm in diameter. (Herein and in the following, all uncertainties are calculated based on counting statistics with a 95% confidence interval (k = 2)). Sodium partitions to the PbS precipitate and also segregates at the matrix/PbS interface. According to Cahn's methodology for the determination of the Gibbsian interfacial excess, two dividing surfaces are defined, which contain the interfacial region but are far enough away from the interface that the matrix concentration is a constant. A reference Na concentration profile, without any interfacial segregation, is calculated using $N_{Na}^0$ as defined by Eq. (13). Tellurium and sulfur are used as reference elements and the concentrations measured close to the dividing surfaces were employed as the concentrations for the matrix and the precipitate.

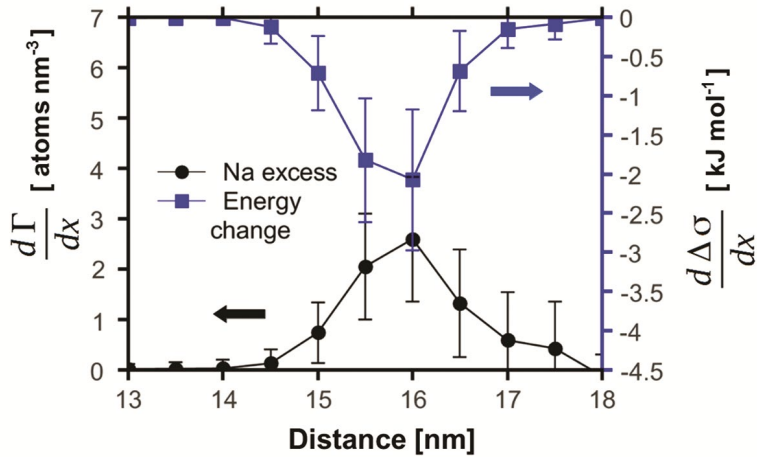

**Figure 3.** Sodium excess concentration profile at the interface between the PbTe matrix and the PbS precipitate and the calculated related decrease in interfacial free energy due to segregation.

The concentration profile of excess Na, dΓ/dx, was calculated based on the difference between the measured concentration profile and the reference concentration profile, Fig. 3. The excess concentration profile, dΓ/dx, is expressed in units of atoms per unit volume of material. The integral of this concentration profile yields the total Gibbsian interfacial excess, which is expressed in number of atoms per unit area of interface. As demonstrated in Ref. (13) the decrease in interfacial free energy depends on the annealing temperature, which isn't precisely defined in our case because precipitation occurs while slowly cooling the sample. One can assume, however, that the observed Na interfacial segregation is in local equilibrium, corresponding to the lowest temperature before Na diffusion is essentially frozen-out on the length scale relevant to our measurements. The diffusion coefficient of Na in PbTe [51], taking into account the effect of Na doping [52], yields a calculated diffusion time of 7 s at 650 K and 40 min at 550 K for a diffusion length of $L = \sqrt{(4Dt)}$ = 10 nm. Therefore, the observed segregation is assumed to correspond to local equilibrium segregation that would be observed at 600 K. This temperature was used in equation (13) to calculate the profile of the decrease in interfacial free energy, $d(\Delta\sigma)/dx$, Fig. 3, which is expressed in kJ per mol of material. Integration of the Na excess concentration profile yields a total Gibbsian interfacial excess of 3.90 ± 0.89 nm$^{-2}$ and integration of the change in interfacial free energy gives a total change of -35 ± 12 mJ m$^{-2}$, which is negative as must be the case.

**4.3. Analysis of a metallic system (Ni-base alloy 600)**

As an example of a metallic system, we studied the commercial Ni-based Alloy 600, which has the composition 16.05Cr-8.8Fe-0.81Mn-0.06C-0.45Si-0.02Cu-0.04Co-0.53(Al + Ti) at.%. As a final heat treatment, the material was annealed at 820 °C for 2 h. During the APT analysis, UV picosecond laser pulses were applied to the APT specimen at a pulse repetition rate of 250 kHz, 0.2 nJ per pulse, and an average detection rate of 0.02 atom per pulse (2 %). APT measurements were performed under ultrahigh vacuum (<2×10$^{-11}$ Torr), with the nanotip at 66.8 K. These conditions yield optimal compositional accuracy for a Ni-Al-Cr alloy [53]. In the case of the Ni-based Alloy 600, carbide precipitates have been observed at grain boundaries for a specific heat-treatment [54,55]. Different segregation behavior patterns have been observed at different interfaces and at the grain boundary, which are likely to have an effect on IGSCC [22,23]. Among the different elements, B segregates significantly at three different interfaces [56,57]. Fig. 4 displays the concentration profiles of the major elements Ni and Cr plus the concentration profile of B across the carbide interface. The B reference concentration profile is calculated using Ni and Cr as the reference elements.

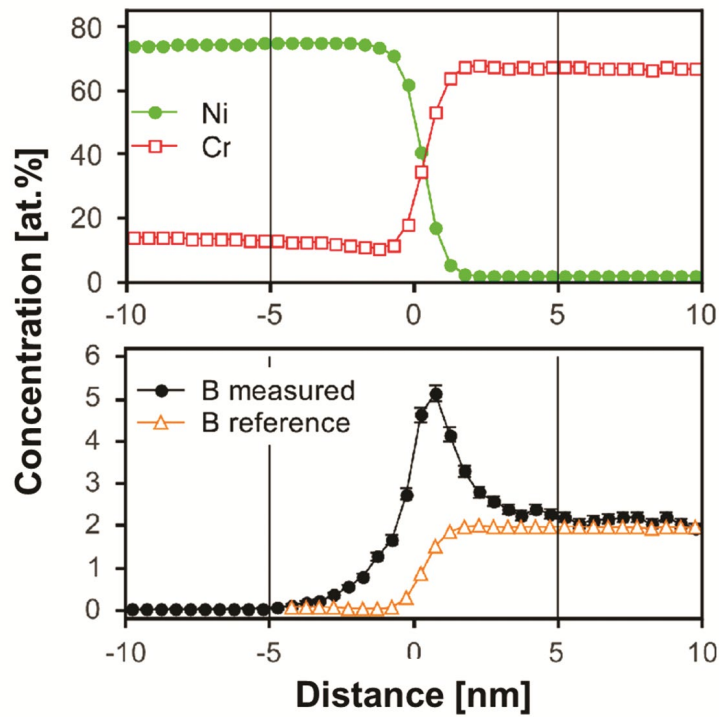

**Figure 4.** Concentration profiles of Ni, Cr and B measured at the heterophase interface between the alloy 600 matrix and a carbide precipitate. The reference B concentration profile is calculated using Ni and Cr as reference elements.

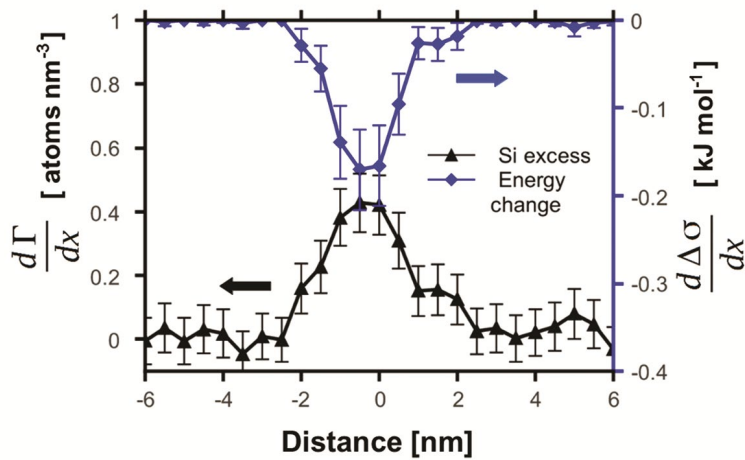

**Figure 5.** Boron excess concentration profile at the interface between the Ni-alloy 600 matrix and a carbide precipitate and the calculated concomitant decrease in interfacial free energy due to segregation.

The resulting B excess concentration profile and decreases in the interfacial free energy are displayed in Fig. 5. The decrease in the interfacial free energy does not have exactly the same shape as the excess concentration profile, illustrating the fact that the decrease in interfacial free energy doesn't only depend on the excess concentration distribution, but also on the reference concentration profile, which is asymmetric because of the partitioning of B to the carbide precipitate. A total integral interfacial excess of 9.8 ± 0.7 nm$^{-2}$ for B was measured with a concomitant integral decrease in the interfacial free energy of -263 ± 42 mJ m$^{-2}$, as calculated using the present method.

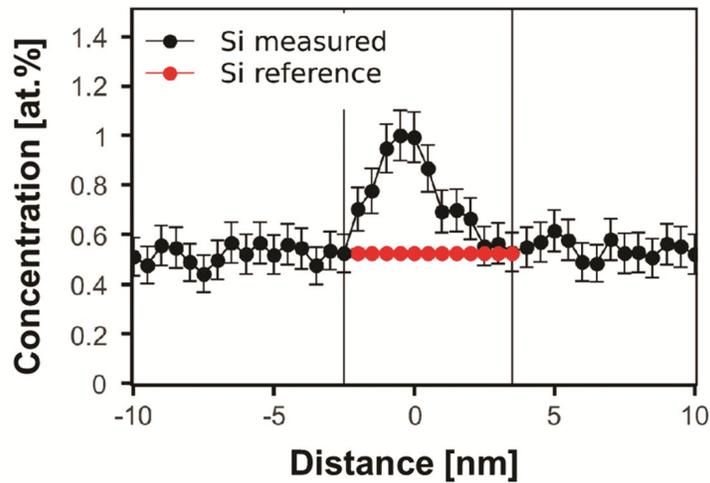

**Figure 6.** Silicon concentration profile measured across a grain boundary in the Ni-alloy 600. The reference B profile is a constant concentration equal to the average concentration of the matrix in the vicinity of the grain boundary.

The same methodology is utilized to characterize Si segregation at a grain boundary (homo-phase interface) in the same alloy, Fig. 6. In this case the Si reference concentration profile is simply a constant concentration, which is equal to the average Si concentration measured in the matrix in the vicinity of the grain boundary. The resulting profile of the interfacial excess and the decrease in the interfacial free energy are displayed in Fig. 7. A total integral interfacial excess of 1.21 ± 0.29 nm$^{-2}$ was measured and, using the methodology presented in this article, a decrease in integral interfacial free energy of -8.9 ± 1.2 mJ.m$^{-2}$ was calculated, which is negative as it must be.

**Table 1.** Summary of the calculated changes in interfacial free energy due to segregation by utilizing the matrix or precipitate solute concentration as reference concentrations for the interface, the average of the two, and by using the continuous method presented in this work.

| Reference | Na in PbTe - PbS | | B in Alloy 600 Carbide | | Si in Alloy 600 GB | |
|---|---|---|---|---|---|---|
| | Concentration [at. %] | $\Delta\sigma$ [mJ m$^{-2}$] | Concentration [at. %] | $\Delta\sigma$ [mJ m$^{-2}$] | Concentration [at. %] | $\Delta\sigma$ [mJ m$^{-2}$] |
| Matrix | 0 | $-\infty$ | 0.018 ± 0.007 | - 834 ± 140 | 0.53 ± 0.02 | - 12.1 ± 3.1 |
| Precipitate | 2.64 ± 0.40 | - 24.2 ± 12 | 1.98 ± 013 | - 141 ± 28 | - | - |
| Average | 1.32 ± 0.20 | - 46 ± 15 | 1.00 ± 0.06 | - 242 ± 40 | - | - |
| Continuous profile | - | - 34.7 ± 11.4 | - | -263 ± 42 | 0.53 ± 0.02 | - 8.88 ± 1.2 |

The different changes in interfacial free energies are summarized in Table 1. For comparison, the Gibbs interfacial free energies were also calculated using Gibbs' formalism, using the matrix or precipitate solute concentration or the average of the two as a reference concentration. *As anticipated, these three methods yield different decreases in the interfacial free energy for each material.* Large or small decreases in interfacial free energy are obtained if one chooses respectively small or large solute concentrations (i.e., matrix or precipitate solute concentrations) as references. For both systems, the results vary by a factor greater than five depending on the specific choice. This illustrates the fact that this method does not have a true thermodynamic meaning. Additionally, for the case of Si segregation at the grain boundary in Alloy 600, the two methods yield different values, even though both use the same average matrix concentration for $c_{initial}$. This is due to the fact that Gibbs' formalism uses the maximum measured concentration for $c_{final}$, whereas the methodology employing a continuous reference concentration profile takes into account the actual shape of the segregation concentration profile at each data point, which is physically more reasonable.

5.  **Discussion**

The new methodology utilized in this research, based on Cahn's approach for internal interfaces, for calculating interfacial excesses, permits the calculation of the decrease in the interfacial free energy due to segregation at heterophase interfaces. It takes into account the actual distribution of solute atoms across an interface, which is absolutely necessary for the calculation of a decrease in the interfacial free energy at internal interfaces. Indeed, when the solute atoms partition to one phase, a given interfacial excess will result in a different free energy decrease whether the atoms are segregating on the solute-rich side or to the solute-poor side of the interfacial concentration profile; specifically, this needs to be taken into account to calculate a physically meaningful decrease in interfacial free energy. Similarly, the exact shape of the concentration profile is very important, but it has been neglected in the traditional method for calculating the decrease in the interfacial free energy, which only takes into account the maximum concentration of an interfacial concentration profile. This is illustrated by the calculation of the decrease in interfacial free energy for Si segregation at a grain boundary in the Ni-alloy 600. Even though there is no partitioning of Si with respect to the interface, the two methods yield different decreases in the interfacial free energy, because only the method using the continuous concentration profile takes account of the actual shape of the segregation profile. For this reason, the continuous method is preferred even in the case of segregation at homophase interfaces (GBs). This is all the more important because homophase boundaries are expected to exhibit complex excess concentration profile shapes, which can even be asymmetric for a symmetric grain boundary [58].

The methodology presented herein is based on assumptions that must be understood when interpreting its quantitative results. The calculation of the interfacial free energy is based on the Gibbs adsorption isotherm equation, which is used for near ideal solutions or solid-solutions that obey Henry's Law with a constant value of the Henry's law constant. Relatively high solute concentrations can, however, be measured at internal interfaces and these approximations may not hold in cases when strong solute-solute atom interactions occur at internal interfaces and/or in the matrix. In this case the method must be adapted to take into account the change of the thermodynamic activity coefficient with concentration. In the absence, however, of accurate data concerning the segregation isotherm one can utilize this methodology to provide the best approximation possible of the decrease in the interfacial free energy. Another source of error may come from errors associated with the measurement of the concentration profile across an internal interface. The spatial resolution of an APT can affect the shape of a concentration profile and, therefore, the characterization of an internal interface needs to be carefully optimized. APT is the only technique currently capable of measuring concentration profiles in a small volume of material and often at the sub-nanometer scale, but it has its own limitations [35,59,60]. It was demonstrated, however, that the most accurate concentration profile across an internal interface is obtained when the interface is perpendicular to a nanotip's long-axis (z-axis) [61–63]; thereby, achieving near atomic resolution perpendicular to an internal interface in some cases [64–66]. Therefore, the most accurate interfacial free energy is measured for internal interfaces oriented perpendicular to a nanotip's z-axis. The interfacial excess of solute atoms can, however, be measured for any interfacial orientation with respect to the z-axis of a nanotip. If well-oriented interfaces are present in the analyzed volume, the present methodology can still be utilized assuming a theoretical or calculated interface shape or a shape measured on another internal interface that is measured with a better spatial resolution and has the same five macroscopic degrees of freedoms [1,14,67]

## 6. Summary

In conclusion, a methodology is developed to calculate the reduction in interfacial free energy caused by segregation at an internal interface:

- In this new methodology, a profile of excess concentration is first calculated using a method based on J.W. Cahn's approach to the Gibbsian interfacial excess. Then the corresponding local reduction in interfacial free energy is calculated using Gibb's adsorption isotherm.

- Detailed applications of our new methodology are presented for: (1) Sodium segregation at a PbTe/PbS interface in a semiconducting thermoelectric system PbTe – PbS – Na (Figs. 2 and 3); (b) Boron segregation at a carbide/metal interface in a Ni-based 600 alloy (Figs. 4 and 5); and (c) Silicon segregation at a homo-phase boundary (GB) in the same alloy (Figs. 6 and 7).

- The methods used in prior articles to calculate a decrease in interfacial free energy require one to define an interface reference solute concentration in the absence of segregation. We demonstrate that the choice made for this reference, the matrix or precipitate solute concentration or the average of the two values, has an important effect on the calculated interfacial free energy, Table 1. We illustrate that our methodology doesn't require this choice and hence it is thermodynamically meaningful. The concentration profile across an interface is utilized to calculate a profile of the decrease in interfacial free energy caused by segregation. In this manner, the actual shape of the internal interface is taken into account.

- In some cases, our methodology may be limited by the assumption of a solution obeying Henry's law or an ideal solution, or by the spatial resolution of the APT. Nevertheless, in the absence of accurate thermodynamic data for the solid solution, our methodology can provide useful thermodynamic information based on the acquired experimental data.

**Acknowledgements**

The atom-probe tomographic measurements were performed at the Northwestern University Center for Atom-Probe Tomography (NUCAPT). The local-electrode atom-probe (LEAP) tomograph was purchased and upgraded with funding from NSF-MRI (DMR-0420532) and ONR-DURIP (N00014-0400798, N00014-0610539, N00014-0910781) grants. This research was supported by the National Science Foundation's MRSEC program (DMR-1121262) and made use of its Shared Facilities at the Materials Research Center of Northwestern University. We also gratefully acknowledge the Initiative for Sustainability and Energy at Northwestern (ISEN) for grants to upgrade the capabilities of NUCAPT. The research on the PbTe-PbS-Na, B system was supported by the US Department of Energy, Office of Science, Office of Basic Energy Sciences, through the Energy Frontier Research Center "Revolutionary Materials for Solid State Energy Conversion," Award Number DE-SC0001054. The research on Alloy 600 was supported by the Office of Basic Energy Sciences, U. S. Department of Energy under contract DE-AC06-76RLO 1830 at Pacific Northwest National Laboratory in cooperation with Drs. S. M. Bruemmer and M. J. Olszta.